\documentclass{aa} % for a referee version

\usepackage[usenames]{color} 
\usepackage[mathcal]{eucal}
\usepackage{amssymb}
\usepackage{amsmath}
\usepackage{natbib}                                   % (traditional abstract)
\usepackage{graphicx}
\usepackage{txfonts}
\newcommand{\dss} {$\delta$ Scuti stars}
\newcommand{\ds} {$\delta$ Scuti}
\newcommand{\gd} {$\gamma$ Dor}
\newcommand{\gds} {$\gamma$ Dor stars}
\newcommand{\msun} {{\rm M}_\odot}
\newcommand{\mmsun} {M/{\rm M}_\odot}

\newcommand{\muHz} {\mu\mbox{Hz}}

\newcommand{\cgs} {\mbox{gcm}^{-3}}

\newcommand{\rhom} {{\bar \rho}}

\newcommand{\teff} {T_{{\rm eff}}}
\newcommand{\amlt} {\alpha_{ML}}
\newcommand{\dov} {d_{\mbox{ov}}}

\newcommand{\lsepl} {\Delta\nu_{\ell}}
\newcommand{\alsepl} {\langle \lsepl \rangle}
\newcommand{\lsep} {\Delta\nu}

\newcommand{\sigrhom} {\sigma(\rhom)}

\newcommand{\rhomt} {{\bar \rho}^{\rm T}}

\newcommand{\sigrhomt} {\sigma(\rhomt)}

\newcommand{\cesam} {{\sc cesam}}

\newcommand{\graco} {{GraCo}}
\newcommand{\most} {\emph{MOST}}
\newcommand{\corot} {\emph{CoRoT}}
\newcommand{\kepler} {\emph{Kepler}}
\newcommand{\plato} {\emph{PLATO}}
\newcommand{\gaia} {\emph{GAIA}}

\newcommand{\tool} {TOUCAN}

\newcommand{\eqn} [1] {
\begin{equation}
#1
\end{equation}}

\begin{document}
%

%   \title{Theoretical properties of mean large spacings in the oscillation spectra of A-F, main-sequence stars}
% \subtitle{VOTA: virtual observatory tool for asteroseismology}
    \title{Measuring mean densities of \dss\ with asteroseismology}  
    \subtitle{Theoretical properties of large separations using \tool}

   \author{J. C. Su\'{a}rez\inst{1} \thanks{e-mail: jcsuarez@iaa.es}
         \and
          A. Garc\'{\i}a Hern\'andez\inst{2,1}
          \and
          A. Moya\inst{3}
         \and
          C. Rodrigo\inst{3,4}
         \and
          E. Solano\inst{3,4}
          \and 
          R. Garrido\inst{1}
          \and
          J.R. Rod\'on\inst{1}}

  \institute{Instituto de Astrof\'{\i}sica de Andaluc\'{\i}a (CSIC), Granada, CP3004 , Spain.
                   \and
                   Centro de Astrof\'{\i}sica, Universidade do Porto, Rua das Estrelas 4150-762, Porto, Portugal.                   
                   \and
                   Dpt. Astrof\'{\i}sica CAB (INTA-CSIC). ESAC Campus. P.O. Box 78. 28691 Villanueva de la Ca\~{n}ada,
                   Madrid, Spain. 
                   \and
	            Spanish Virtual Observatory.}

   \date{Received ; accepted}

% \abstract{}{}{}{}{}
% 5 {} token are mandatory
%--------------------------------------------------------------------------------------------------
  \abstract
%--------------------------------------------------------------------------------------------------
  % context heading (optional)
  % {} leave it empty if necessary
   {}
  % aims heading (mandatory)
   {We aim at studying the theoretical properties of the regular spacings found in the oscillation spectra
   of  \dss. }
  % methods heading (mandatory)
   {We performed a multi-variable analysis covering  a wide range of stellar structure and seismic properties and
    model parameters representative of intermediate-mass, main sequence stars. The work-flow is
    entirely done using a new Virtual Observatory tool: \tool\ (the VO gateway for asteroseismic models), which is presented in this paper. }
  % results heading (mandatory)
  {A linear relation between the large separation and the mean density is predicted to be found in the 
    low frequency frequency domain (i.e. radial orders spanning from 1 to 8, approximately) of the 
    main-sequence, delta Scuti  stars' oscillation spectrum.
   We found that such a linear behavior stands whatever the mass, metallicity, mixing length, and 
   overshooting parameters considered in this work. The intrinsic error of the method is
   discussed. This includes the uncertainty in the large separation determination and the role of rotation.
   The validity of the relation found is only guaranteed for stars rotating up to 40\% of their break-up
   velocity. 
   Finally, we applied the diagnostic method presented in this work to five stars for which regular patterns have been 
   found. Our estimates for the mean density and the frequency of the fundamental radial mode  
   match with those given in the literature within a 20\% of deviation.}  
  % conclusions heading (optional), leave it empty if necessary
   {Asteroseismology has thus revealed an independent direct measure of the average density of \dss, analogous 
   to that of the Sun.  
    This places tight constraints on the mode identification and hence on the stellar internal 
   structure and dynamics, and allows a determination the radius of planets orbiting around \dss\ with unprecedented 
   precision. This opens the way for studying the evolution of regular patterns in pulsating stars, and its
   relation with stellar structure and evolution.}

   \keywords{asteroseismology, stars: evolution, stars:variables: delta scuti,
             stars: oscillations (including pulsations), stars: interiors, virtual observatory tools}

   \maketitle
%--------------------------------------------------------------------------------------------------
%--------------------------------------------------------------------------------------------------
%--------------------------------------------------------------------------------------------------

%--------------------------------------------------------------------------------------------------
\section{Introduction\label{sec:intro}}
%--------------------------------------------------------------------------------------------------
 
Nowadays, thanks to the asteroseismic space missions like \most\ \citep{most03}, \corot\ \citep{Corot03}, and
\kepler\ \citep{Gilliland10}, the study of the intrinsic variability of A- and F-type stars is living a revolution.
In particular, the large number of modes detected and the variety of frequency domain covered has
given rise to the so-called \emph{hybrid} phenomenon, which imposes a revision of the current
observational instability strips of \ds\ and \gd\ stars \citep[see ][ and references therein]{Uytt11}. 
Those stars have been found to be located over the entire \gd\ and \ds\ instability strips, which implies that 
a review of their pulsation mechanisms is necessary to supplement (or even substitute) the convective blocking
effect and the $\kappa$ mechanisms, respectively.  From the theoretical side, based on stellar energy
balance studies, \citet{MoyaCris10} concluded that \dss\ are able to excite hundreds of pulsation modes,
and whose accumulated pulsation energy is not large enough to destroy their hydrostatic equilibrium. 
Likewise, the richness found in the oscillation spectrum
of these stars is interpreted by \citet{Kallinger10} as non-radial pulsation superimposed on granulation noise
(correlated noise).
 
In this work we focus on \dss\ which are intermediate-mass (i.e., from 1.4 to 3$\,\msun$, approximately), 
mainly in the main sequence and the sub-giant branch, but also in the pre-main sequence. Their spectral 
types range from A2 to F5, and their luminosity classes go from IV to V, approximately. This locates them at 
the lower part of the Cepheid's instability strip. Their pulsations are mainly driven by the so-called $\kappa$ 
mechanism \citep[see reviews by ][]{Breger00rev, Handler09rev}, showing radial and non-radial oscillation 
modes excited in a frequency domain ranging from a few tens of $\muHz$ up to about $800-900\,\muHz$ 
\citep[see e.g. ][]{Poretti11}.

The interpretation of the oscillation spectra  of these stars has never been an easy task. Due to the complexity of their
oscillation spectra the identification of detected modes is often difficult. A unique mode identification is often 
impossible and this hampers the seismology studies for these stars. Additional uncertainties arise from the effect 
of rapid rotation, both directly on the hydrostatic balance in the star and, perhaps more importantly, through 
mixing caused by circulation or instabilities induced by rotation \citep[see e.g. ][]{Zahn92}.

However some decades ago, this scenario started to change thanks to the detection of regular patterns in
the detected frequencies. These patterns were observed in the low radial order modes (mixed modes)
frequency domain, in which main-sequence classical pulsators show the maximum of their oscillation 
power,  i.e. between 80 and 800 $\muHz$, hereafter intermediate frequency domain. There,
regularities similar to those found in solar-like stars 
were not expected \citep[for a review on this topic, see][]{Goupil05}. Thanks to great efforts made in 
long ground-based multi-site campaigns, \cite{Handler97} found regularities
(near $26\,\muHz$) in the oscillation spectrum of the young \ds\ star \object{CD-24 7599}, 
which contained 13 frequencies. Those regularities were found using a Fourier transform technique. 
Also \cite{Breger99} found a regular spacing close to $46\,\muHz$ in the oscillation spectrum of the
\ds\ star \object{FG\,Vir}, composed of 24 frequencies, searching
for  regular spacings using histograms of frequency differences, instead of Fourier transform.
Later on, they re-studied the star using data spanned in ten years \citep{Breger09}, in which 
the number of detected frequencies increased to 68, approximately.
The frequency spacing found was attributed to an observed clustering of certain non-radial
modes around radial ones. This clustering was explained by the higher probability of
certain $\ell=1$ modes \citep[the so-called trapped modes ][]{Dziembowski90}  to be excited 
to observable amplitudes than other modes.

Thanks to the different space missions providing asteroseismic data, similar studies were
undertaken, with much more frequencies detected with very high precision. In 2007, 
\citet{Matt07} found 88 frequencies in the oscillation spectrum of the \ds\ star 
\object{HD209775} observed by \most. In that work, a regular spacing of $50\,\muHz$ 
approximately was obtained using histograms of frequency differences. This was 
assumed to be a large separation, although no physical explanation to support this was
given. Recently, regular patterns were also found in the oscillation spectra of  \dss\ 
observed by \corot\  \citep{GH09,Manteg12, GH13aa} and \kepler\ \citep{GH13sp}, 
and the hypothesis of identifying them with the large separation has been 
reinforced. 

Theoretically large separation is expected to grow from low to high 
radial orders. In contrast to the well-defined plateau shown by the large separation
in the high radial order domain 
\citep[the so-called asymptotic regime, see e.g. ][]{Antoci11, Zwintz11}, it forms
a quasi-periodic structure at low radial order regime \citep[see Fig.~4 of ][]{GH09}.
In that work, it was shown that the standard deviation of such an structure is
roughly $2.5\,\muHz$, and the distance between the mean value of this structure and 
the high order regime  is of the order of $5\,\muHz$.

The main questions we intend to answer in the present work are: What are the physical properties 
of the periodicities observed in the low frequency domain of \dss? Do 
they have similar properties to those found in solar-like stars? 

The above questions are tackled by analyzing the properties of the predicted regularities over
a large collection of models and physical variables. To do so, we developed a virtual observatory
tool, \tool\footnote{http://svo.cab.inta-csic.es/theory/sisms3/index.php?}, 
designed to easily handle stellar and seismic
models, examine their properties, compare them with observational data and find models
representative of the studied star(s). The tool is presented in this paper.

%--------------------------------------------------------------------------------------------------
\section{The method \label{sec:method}}
%--------------------------------------------------------------------------------------------------

We examine a dense sample of asteroseismic models 
representative of A-F main-sequence stars, i.e. covering the corresponding area in the HR diagram where 
classical pulsations for these stars are expected. Such a modeling approach requires an efficient computing
procedure as well as the capability of managing and analyzing large sets of models. 
For the former, we performed most of the computation using the GRID computing service
provided at IAA-CSIC as one of the nodes of the \emph{Ibergrid} virtual organization.

For the latter, we have developed \tool, a VO tool able to easily compare different and heterogeneous collections of 
asteroseismic models (equilibrium models and their corresponding synthetic oscillation spectra). Details about
the work-flow followed and/or the different \tool\ services can be found in Appendix~\ref{apsec:workflow}
and \ref{apsec:voservice}, respectively. The model collection used in the present work is
described in the next section.

%-tab-%
\begin{table}
  \begin{center}
    \caption{Ranges of the four parameters used to construct the current model dataset
             representative of intermediate-mass stars.}
    \vspace{1em}
    \renewcommand{\arraystretch}{1.2}
    \begin{tabular}[h]{cccc}
      \hline
        Parameter & Lowest  & Highest & Step \\
      \hline
       $\mmsun$  &  1.25   &  2.20 & 0.01  \\
       $[Fe/H]$   & -0.52   & +0.08 & 0.20  \\
       $\amlt$    &  0.50   &  1.50 & 0.50  \\
       $\dov$     &  0.10   &  0.30 & 0.10  \\
      \hline
      \end{tabular}
    \label{tab:param}
  \end{center}
\end{table}
%-tab-%

%--------------------------------------------------------------------------------------------------
\subsection{The grid of models\label{ssec:votadb}}
%--------------------------------------------------------------------------------------------------

We constructed a model collection composed of approximately $5\,10^5$ models
of intermediate-mass stars (namely \ds\ and \gds). For the sake of homogeneity 
and precision of the asteroseismic mode sample, models were computed following the prescriptions 
suggested by ESTA/CoRoT\footnote{http://www.astro.up.pt/corot/} 
 working group \citep{Moya08esta, Lebreton08esta}. 

The equilibrium models were computed with the evolutionary code \cesam\ \citep{Morel97, MorelLeb08}.
Oscillation frequencies were computed with \graco\ \citep{Moya04, Moya08graco}, which uses the perturbative approach, 
to provide adiabatic and non-adiabatic quantities related to pulsation and includes the convection-pulsation 
interaction using the Time Dependent Convection theory \citep[TDC, ][]{Dupret05cpc2}. 

The model grid is composed of evolutionary tracks, evolved 
from ZAMS up to the sub-giant branch. Each track contains about 250 equilibrium models with their 
corresponding oscillation spectra.  The equilibrium models were computed varying four quantities: two 
global stellar parameters, mass and metallicity ($M$ and [Fe/H]), and two modeling parameters often used 
in asteroseismology, the convection efficiency, $\amlt=l/H_{\rm p}$,  where $l$ is the mixing length and 
$H_{\rm p}$ is the pressure scale height, and the overshoot parameter $\dov=l_{\rm ov}/H_{\rm p}$ ($l_{\rm ov}$
being the penetration length of the convective elements). The range of variation for each
parameter in this dataset is listed in Table~\ref{tab:param}. The physical parameters of the models
were chosen as the typically adopted for \dss.

Although \dss\ have also been observed in the pre and post-main sequence evolutionary stages,
the present work focuses on the main sequence (i.e. hydrogen-burning phase in the
convective core) where the low-order periodicity has been found (see previous section).

%fig%
\begin{figure*}[ht!]
 \begin{center}
  \includegraphics[width=8.5cm]{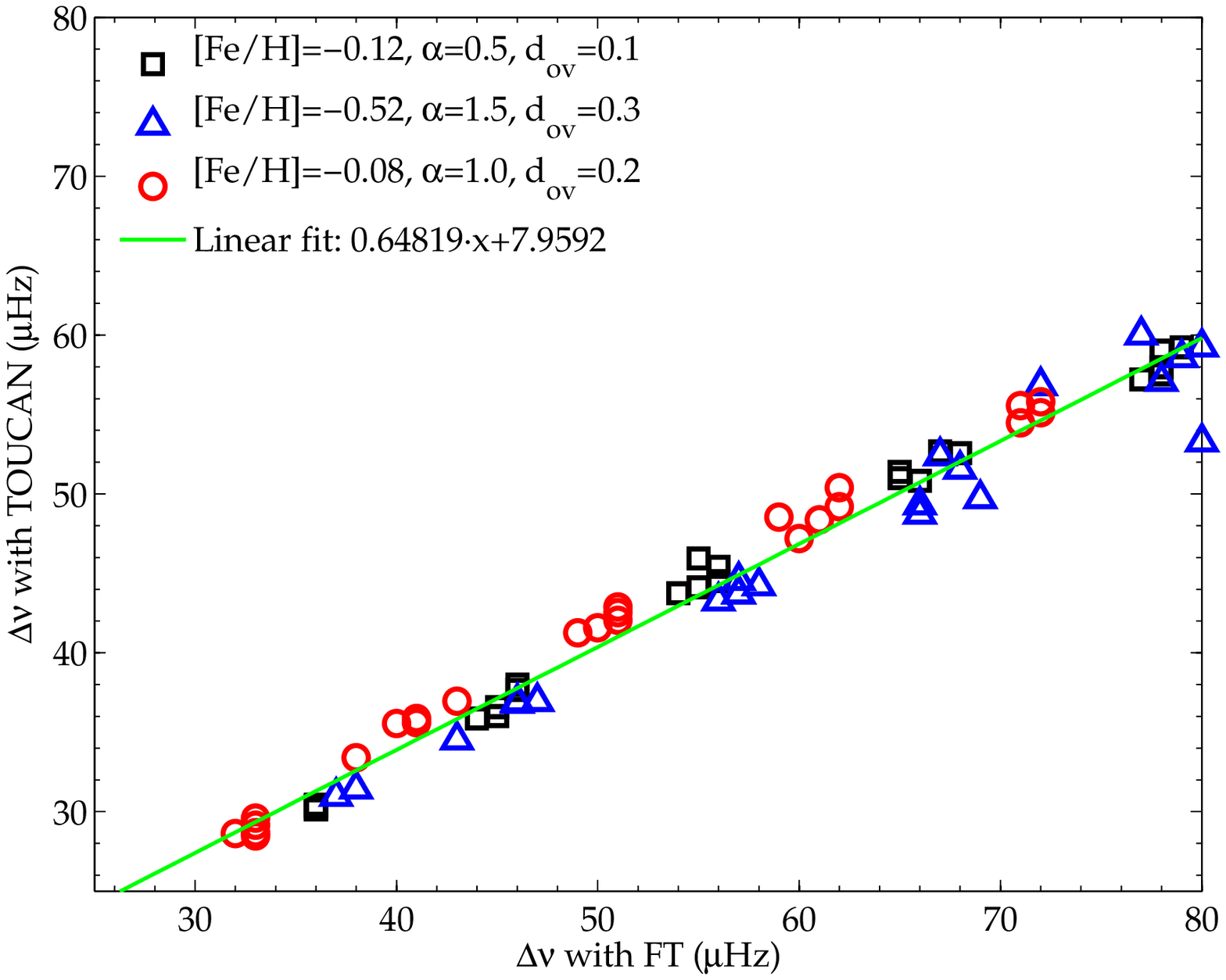}    
  \includegraphics[width=8.5cm]{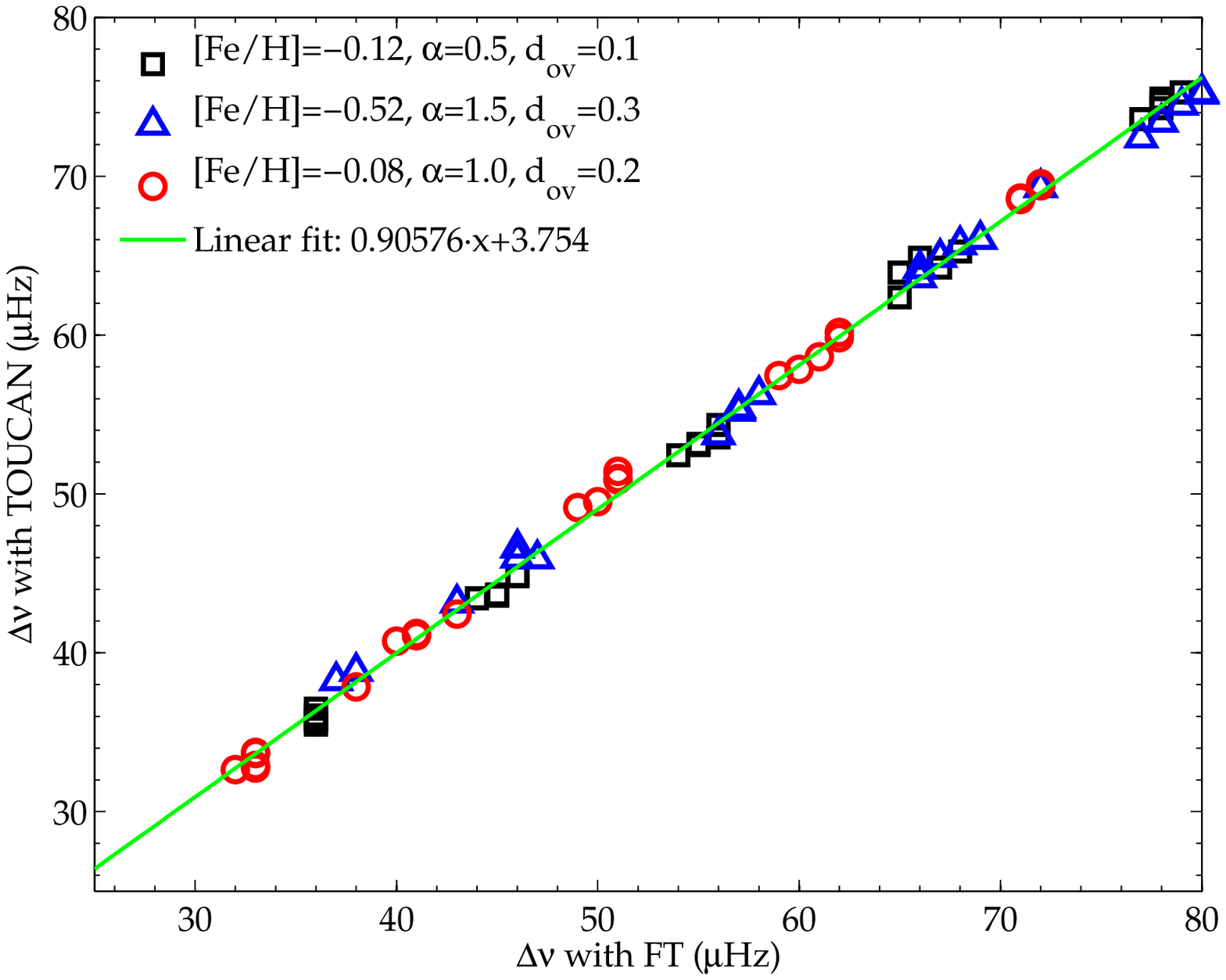}
  \caption{Left panel: spacings found using the FT method are compared to large separation 
                 values computed with \tool\ in the range of the observed modes for \dss. 
                 Right panel: Similar to left panel, but only p modes, with
                 radial order $n\ge1$ were considered in the calculation of \tool\
                 $\lsep$. Spacings obtained with the FT method and \tool\ computations are 
                 only equivalent under the conditions adopted in
                 the right panel. Colors are only available in the online version of the paper.}
  \label{fig:Dnu_FTvsDnu_VO}
 \end{center}
\end{figure*}
%fig%

%--------------------------------------------------------------------------------------------------
\subsection{Large frequency separation \label{ssec:ls}}
%--------------------------------------------------------------------------------------------------

The frequency difference defined as 
\eqn{\lsep_\ell=\nu_{n,\ell}-\nu_{n-1,\ell}\label{eq:def_lsep}}
is known as a large separation, where $\nu_{n,\ell}$ is the frequency of a mode with radial order 
$n$ and degree $\ell$, whatever the azimuthal order $m$. This frequency difference is nearly 
constant for $p$ modes in the asymptotic regime, and obeys, in a first approximation, the following 
physical dependence \citep{Tassoul80,Gough90}:
\eqn{\lsep=2\Bigg[\int_{0}^{R}\frac{\mbox{dr}}{c_s}\Bigg]^{-1}=
     {\tau}^{-1}\label{eq:lsep_dep}}
where $c_{s}$ is the sound speed and ${\tau}$ is the acoustic time between 0 and $R$ 
(i.e. the stellar center and surface, respectively) which define the stellar region in which 
the oscillation mode is propagating. 

\tool\ calculates $\lsep$ everywhere in the theoretical oscillation spectrum, for any combination 
of angular degree (up to $\ell=3$), frequency and radial order domains. In practice, the calculation
is performed in two steps. First, for a given angular degree
$\ell$, i.e.
\eqn{\lsepl = \frac{1}{N_{n}}\sum_{n=n_j}^{n_{k-1}} (\nu_{n+1}-\nu_n),~~~~j<k
\label{eq:votadef_lsep}}
where $n$ is the radial order, and $N_n=n_{k}-n_j$ is the number of frequencies 
found in the interval [$n_j,n_{k}$] for the given $\ell$.
Then \tool\ calculates the average of $\Delta\nu_{\ell}$ over all the angular degrees 
calculated in the model, i.e.,
\eqn{\lsep=\alsepl = \frac{1}{N_{\ell}}\sum_{\ell=\ell_i}^{\ell_j} \lsepl \label{eq:votadef_alsep}}
where $N_\ell=\ell_j-\ell_i$ is the total number of angular degrees considered. 
Here we are interested in studying the physical properties of such separations 
for the intermediate frequency domain. But, before analysing the \tool\ results, 
we need to take into account some considerations
with respect to the rotation and the presence of mixed modes.

%--------------------------------------------------------------------------------------------------
\subsection{Rotation and mixed modes\label{ssec:rolerot}}
%--------------------------------------------------------------------------------------------------

The A-F type stars are generally fast rotators, and rotation effects on both the stellar
structure and the oscillation frequencies cannot be neglected, particularly those 
caused by stellar distortion \citep{Soufi98,Sua06rotcel}, even for $m=0$ modes 
with which $\lsep$ are calculated in the present work \citep{Sua10apj}.

Rotation effects on oscillations is commonly taken into account through the 
perturbation approximation \citep{DG92}, which is limited to slow-to-moderate
rotation, i.e. small stellar deformations \citep[see e.g. ][ for semi-empirical and theoretical
studies of the limitations of the perturbation approximation]{Sua05altairII, Reese06}.
Therefore, for moderate-to-rapid rotators a non-pertubative approach for the calculation 
of the oscillation modes on a deformed star becomes necessary \citep[e.g. ][]{Lig06}. 
However, nowadays this calculation is available for polytropic models and for some 
more realistic fully deformed 2D stellar models on the ZAMS based on the self-consistent 
field (SCF) method \citep{Jackson05,McGregor07, Reese09} which considers the models
chemically homogenous with angular velocity assumed to be dependent only on the 
distance from the axis of rotation. Furthermore, these latter models together with the calculation
of non-perturbative oscillations require a significant amount of computing resources
as well as time of computation. Therefore the use of a proper modeling for 
rapidly rotating stars would be unpractical for the present work.

On the other hand, since periodicities are indeed observed, 
it might be concluded that rotation effects are not sufficient to break the regularities.
Indeed, non-perturbative calculations of the oscillation spectra for rapidly rotating 
polytropic models indicate that as rotation increases, the asymptotic structure of 
the non-rotating frequency spectrum is replaced by a new form of organization
\citep{Reese08, Reese09}. This new mode frequency 
organization also exhibits regular structures, including the large separation  
\citep[see ][ for a recent theoretical work on regular patterns in rapidly-rotating stars]{Lig10an}
whose variation (normalized by the density) from the non-rotating case is negligible \citep{Lig06}.
Furthermore, calculations of non-perturbative oscillation frequencies on 
SCF models shows a maximum variation of the large separation of around $2.3\,\muHz$ for stars rotating 
up to 40\% of the keplerian velocity (Reese, private communication), which is small compared 
with the precision with which the periodicities are predicted at both low- and high-frequency domains
\citep[$10\muHz$ approx., see e.g. ][]{Manteg12}. 
Considering all the above theoretical arguments, we are allowed to use non-rotating
models for the present study. 

This thus reinforces the hypothesis that identifies the observed periodicities with 
the large separation, as well as the hypothesis that these appear in a new distribution of modes 
predicted by the non-perturbative theory. In order to check those hypothesis,
it is necessary to analyze the oscillation spectra of many stars with different rotational 
velocities. Thanks to space missions like \corot, \kepler, or \plato, this study
can be tackled in the near future.

Another issue that might hamper the detection of periodicities, in particular, of
the large separation, is the presence of mixed modes. This phenomenon is implicitly
considered in this work since the selected models cover all the main sequence. 
In contrast, this cannot be properly studied within the non-perturbative approach, 
since both the polytropic and the SCF models in ZAMS are not expected to properly 
show the avoided crossing 
phenomenon.  In order to properly understand how avoided-crossing may affect the 
detection of periodicities, the present analysis should be done for a more evolved 
SCF model (work in progress). 

On the other hand, all these results were obtained assuming no dependence
with the visibility of the modes which may introduce artificial periodicities that
may potentially be confused with the large separation. Recently, this has been studied 
in the framework of the non-perturbative theory. Specifically, it has been found 
\citep{Reese13} that acoustic modes with the same ($\ell$, $|m|$) values tend 
to have similar amplitude ratios, although this effect is not systematic.  
Since the global visibility of the modes decreases while the mode degree increases,
we do not expect this effect to affect the large separation determination
significantly. In any case, a work on the influence of mode visibilities in the 
study of periodicities for rapid rotators is currently ongoing.

%--------------------------------------------------------------------------------------------------
\subsection{Large separations computation \label{sec:ls-dss}}
%--------------------------------------------------------------------------------------------------
In GH09, regular spacings were found using the Fourier transform (hereafter, FT). 
Here we study whether this technique and the average of the
periodicity used by \tool\ are equivalent.

To do so, we used a subset of asteroseismic models within typical values for $\teff$
and $\log g$ of \dss\ with their corresponding uncertainties. 
For completeness, we considered the following extreme combinations of the 
physical parameters listed in Table~\ref{tab:param} divided into three sets: 
\begin{itemize}
	\item[\#1] [Fe/H]=-0.12, $\amlt=0.5$, and $\dov=0.1$
	\item[\#2] [Fe/H]=-0.52, $\amlt=1.5$, and $\dov=0.3$
	\item[\#3] [Fe/H]=0.08, $\amlt=1$, and $\dov=0.2$
\end{itemize}
For each set of parameters, we selected five models covering the extremes values (four 
models) and the central value (one model) of effective temperature and gravity 
considered in the selected model subset.

For each set of five models, we computed the large separation using both methods in the 
intermediate frequency domain, i.e., no restrictions in the radial orders but 
only in the frequency range were considered in order to mimic the observations.
Independently of the combination of model parameters 
the relation between the large separations obtained with the FT method and those obtained 
by \tool\ was found to be linear (Fig.~\ref{fig:Dnu_FTvsDnu_VO}, left panel). 

On the other hand, the corresponding linear fit does not have a slope equal to one, 
which would be expected if the two methods were equivalent.
In order to understand this, we first attempted to explain such discrepancies by the 
presence of gravity modes (g modes).  However, when removing g modes in \tool\ 
computations (not in the FT ones), the slope of the trend was improved but not equal 
to one, neither.
%fig%
\begin{figure*}[!ht]
   \centering
      \includegraphics[width=13cm]{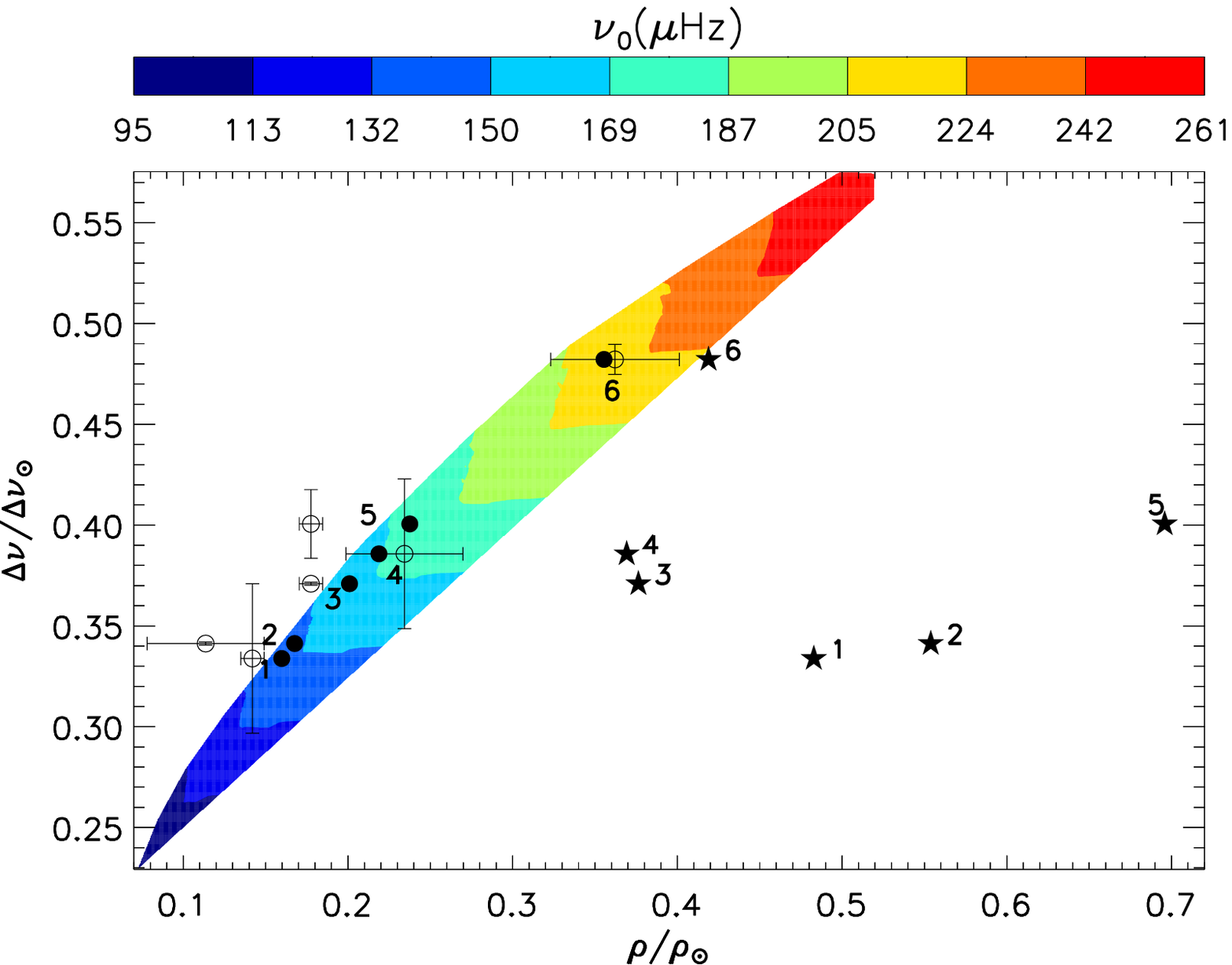}    
          \caption{Predicted large separation as a function of the mean density of the star, normalized 
          		to their solar values, $\lsep_\odot = 134.8 \muHz$ \citep{Kjeldsen08} and 
			$\rhom_\odot= 1.48 {\rm gcm}^{-3}$ \citep{HandbookCP12}, respectively. 
			Contours in gray scale indicate the predicted 
			frequency of the fundamental radial mode. Filled dots,  empty dots, and star symbols
			represent mean densities found in this work, in the literature, and using Tingley's
			calibration, respectively. Stars are labelled as in Table~\ref{tab:obs}.			
			 For the sake of clarity, the error bars in 
			 star symbol estimates are omitted, since they are larger than the abscissa range.
			 Colors are only available in the online version of the paper.} 
     \label{fig:figcontour}
\end{figure*}
%fig%
We then considered the well-known ratio between the period of the fundamental radial 
mode and the period of its first overtone (which is 0.77 for instance, for population I,
main sequence stars). This ratio, which is constant for each \ds\ star, implies a relation
between these two radial modes which does not necessarily follows the average 
behavior of $\lsep$. To avoid such a contribution in the calculation of the large 
spacings by \tool, we also restricted the oscillation modes to those with a radial order
$n\ge1$. 

With all these requirements, the FT and \tool\ large separations are found to be 
formally equivalent (Fig.~\ref{fig:Dnu_FTvsDnu_VO}, right panel). These requisites were thus 
adopted throughout  the whole work. 

Note that we always used the same range of frequencies for computing the FT. This test emphasizes the 
robustness of the FT method to detect the large separation even when g modes and the fundamental radial mode 
are not discriminated. Detailed analysis of the slopes obtained in the two cases indicates that an uncertainty of 
10\% in the equivalence should be considered if other constraints, namely in the frequency range, are not taken 
into account.

%--------------------------------------------------------------------------------------------------
\section{Large separation vs. mean density \label{ssec:ls-vs-ro}}
%--------------------------------------------------------------------------------------------------

One of the main characteristic of large separations observed at high frequency
(e.g. in solar-like stars) is that they are proportional to the stellar mean density 
through Eq.~\ref{eq:lsep_dep}, which, together with another regularity,
the so-called small separation, provides a direct measure of the mass, radius, and evolutionary
stage of those stars \citep[see e.g., ][]{JCD88}. It is thus worthy to investigate if such a 
dependence is also predicted at low frequencies, in particular in the frequency
domain where \dss\ pulsate.

To do so, we followed the workflow described in Appendix~\ref{apsec:workflow} to select 
a set of models to work with. In particular, we considered all the models contained in the model 
database described in the previous sections, i.e. those whose parameters are described in 
Table~\ref{tab:param}. For the sake of simplicity, the study is restricted to models in the main 
sequence, that is, between zero-age to turn-off main sequence stages. In what regards
the asteroseismic content, we adopted the prescriptions deduced in the previous section, i.e., 
we computed the large separation in the intermediate frequency domain using radial and non-radial 
modes with $n\ge1$, and $\ell\le3$. Indeed, the presence of modes with $\ell>3$, whose
presence is suggested by spectroscopic studies \citep{Manteg12},  would 
hamper the determination of the large separation.  Since we are dealing with
oscillation spectra obtained from photometry, and since the FT method (details in GH09) uses the modes 
with the largest amplitudes, it is plausible to consider that those modes correspond to $\ell=[0,3]$ modes.

Analysis of the large separation predicted in the intermediate frequency domain as a function of the mean 
density reveals (Fig.~\ref{fig:figcontour}) a clear relation which can be expressed mathematically as 
\eqn{\lsep/\lsep_\odot = 0.776\,\big(\rho/\rho_\odot\big)^{\,0.46}\,.\label{eq:linreg}}
obtained by performing a linear regression of the complete set of models. 
This gives us an approximation of the overall behavior of the relation 
between $\lsep$ and $\rhom$ in the intermediate frequency domain which is predicted here to be proportional
to the mean density of the star to the power 0.46, which is quite close
to $\lsep\propto\rho^{1/2}$, predicted for high radial orders. This opens the way for studying 
the evolution of regular patterns in pulsating stars throughout the HR diagram, and its relation with stellar evolution.
This also might provide new insights in the hybrid phenomenon, i.e. the yet unexplained 
excitation modes in a wide frequency range covering typically the $\gamma$ Doradus,
$\delta$ Scuti and even solar-like oscillation modes \citep{Uytt11}.

We present no errors in the coefficients of the fitting because none of the theoretical 
$(\lsep_i, \rhom_i)$ are independent from each other, and therefore regression error estimates 
are meaningless. Nevertheless it is possible to examine the domain of validity of the
method. In Appendix~\ref{apsec:errors} we provide estimates of intrinsic errors
of the method and their relation with the domain of validity of the present method.

These results imply that Fig.~\ref{fig:figcontour} can be used
as a powerful diagnostic tool for the study of A-F stars, like \dss\ and/or hybrid
stars which are being observed by the space missions. Not only it provides a direct
estimate of the mean density of the stars,  but also an estimate of the frequency of
the fundamental radial mode. Up to date, this latter asteroseismic observable has not 
been fully exploited in \dss\ 
\citep[with the exception of HADS, see e.g. ][]{Poretti05hads, Sua07pdrotII}
because of the well-known difficulties for the mode identification. Therefore 
the present results also represents an additional help for the mode identification for
this type of pulsator.

The strength of this diagnostic tool is that it is almost model independent, since all
the models contained in the heterogenous dataset follow the same trend.  
In the following sections, we discuss different details of this finding and its consequences.
%-tab-%
\begin{table*}
  \begin{center}
    \caption{Estimates of mean density and frequency of the fundamental radial mode of
    a list of \dss\ found ound in the literature for which regularities in their powerspectra
    were associated with a large spacing. These estimates are compared with ours using 
    the diagnostics presented in this work.
    From the left to right, columns represent the star's identification, 
      the observed $\lsep$, its uncertainty $u$, determinations of the mean density $\rhom$ with their 
      corresponding errors $\sigrhom$, and determinations of the fundamental radial mode frequency. 
      Quantities in parentheses were calculated in this work. The last three columns give the
     estimates of the mean density of each star using the Tingley's method, together with their
     errors, and average $J-K$ magnitude differences used for its calibration.}
    \vspace{1em}
    \renewcommand{\arraystretch}{1.2}
    \begin{tabular}[h]{llcccccccc}
      \hline
           No.		&	Star                  & $\lsep$            & $u$            & $\rhom$    & $\sigrhom$  & $\nu_0$  & $\rhomt$ & $\sigrhomt$ & $<J-K>$\\
                                & $(\muHz)$        & $(\muHz)$   & $(\cgs)$       & $(\cgs)$       & $(\muHz)$ & $(\cgs)$     & $(\cgs)$       & ({\rm mag})\\
      \hline
	 1	&	${\rm HD\,50870}^6$     & $45^{\rm b}$  & 5       & - (0.20)      & - (0.17)       &  80.09(162.352)  & 0.68 & 2.55 & 0.14\\
	 2	&	${\rm FG\,Vir}^2$           & $46^{\rm a}$  & ...        & 0.16(0.24) & 0.10(-)        & 140.62(157.05)  & 0.78 & 2.54 & 0.17\\
	 3	&	${\rm HD\,209775}^5$  & $50^{\rm a}$  & ...       & 0.25 (0.29)& - (0.17)       &  ... (168.55)          & 0.53 & 2.56 & 0.09\\
	 4	&	${\rm HD\,174936}^3$   & $52^{\rm b}$  & 5       & 0.33(0.31) & 0.10(0.17)  & ... (174.72)           & 0.52 & 2.56 & 0.09\\		
	 5	&	${\rm XX\,Pyx}^4$          & $54^{\rm b}$  & 2.3    & 0.25(0.33) & 0.02(0.12)  & ... (180.87)           & 0.98 & 2.55 & 0.23\\
	 6	&	${\rm HD\,174966}^1$  & $65^{\rm b}$  & 1       & 0.51(0.48) & 0.11(0.17)  &  200.23 (212.14) & 0.59 & 2.56 & 0.11 \\
	\hline
      \end{tabular}
      \tablebib{(a)~Histograms of frequency differences; (b)~Fourier Transform of the oscillation powerspectrum;
		     (1)~\citet{GH13aa}; (2)~\citet{Breger09}; (3)~GH09; (4)~\citet{Handler97}; (5)~\citet{Matt07}; 
		     (6) \citet{Manteg12}.}
    \label{tab:obs}
  \end{center}
\end{table*}
%-tab-%

%--------------------------------------------------------------------------------------------------
\section{Some real examples\label{ssec:examples}}
%--------------------------------------------------------------------------------------------------

In order to have a first rough quality check of the diagnostic tool here presented, we applied it 
to some known \dss: \object{FG Vir}, \object{HD 174936}, \object{HD 174966}, \object{XX Pyx},
\object{HD 50870}, and \object{HD 209775} (see references in Table~\ref{tab:obs}), for which the 
regular patterns have been found and analyzed. The comparison of the results found using 
Fig.~\ref{fig:figcontour} and/or Eq.~\ref{eq:linreg} with those published in the literature are is given in Table~\ref{tab:obs}.

In general, predicted values are found within the uncertainties provided in the literature. 
As an example, a value of $52\,\muHz$ was proposed as possible large spacing for the 
\corot\ \ds\ star \object{HD 174936} by GH09. For such a value of the large spacing, and assuming no
error on this value, a mean density of about $\rho=0.31\,{\rm gcm}^{-3}$ is predicted in the present 
work. Our results are in good agreement with those of GH09, taking into account that an uncertainty in 
$\lsep$ of about $5\,\muHz$  was considered in the latter.

We compared the above results with an independent method for estimating the mean density
of stars without making use of neither models nor asteroseismic information. 
More specifically we use the method that uses a color-density calibration for main sequence stars
\citep[see details in ][]{Tingley11}.
This calibration is independent of factors such as age and metallicity, although it may suffer 
from significant error if extinction is unknown or poorly estimated. The  method applied
to our selected stars yields the values given in Table~\ref{tab:obs}. The uncertainties
in the density measurements are significantly larger than those obtained by asteroseismology
(works cited in Table~\ref{tab:obs}) and those obtained in the present work. This prevent
us to conclude anything about a possible trend between the density and the large separation.
 
In any case, to date, the number of A-F stars for which an in-depth study of their regularities
have been performed is very small. Fortunately, thanks to the asteroseismic
space missions, the number of observed A-F stars with high precision increases day by day.
With this diagnostic tool it will be possible to better study asteroseismic properties of 
A-F stars, including the large spacing itself. 

%--------------------------------------------------------------------------------------------------
\section{Conclusions \label{sec:conclusions}}
%--------------------------------------------------------------------------------------------------

We have studied the theoretical properties of the regular spacings 
in the oscillation spectra of \dss. A comprehensive dataset of models representative of A-F 
stars covering a wide range of stellar physical magnitudes (e.g. effective temperature,
gravity, metallicity, etc.)  and model parameters has been 
performed.  We have analyzed the behavior of the predicted large spacings (calculated in the 
intermediate frequency domain, i.e. the frequency domain
in which \dss\ pulsate) and their possible dependencies upon other physical magnitudes 
and/or model parameters. The work-flow has entirely been done using \tool, a virtual observatory tool
for managing and analyzing asteroseismic models that we have developed within the Spanish 
Virtual Observatory (see Appendix~\ref{apsec:tool}).

Firstly, we have shown that,  for the frequency domain where \dss\ pulsate, the regular 
spacings obtained using the FT technique (as in GH09) and the regular mathematical definition 
of large separation computed by \tool, are only equivalent when selecting oscillation modes with 
radial order $n\ge1$.

Secondly, we have found a linear relation between large spacing and
the mean density (in a logarithmic scale), for models in the main sequence, whatever
the metallicity, mass or convection parameters ($\amlt$ and $\dov$) considered (within
the values typically used for \dss).  With the use of a large model grid, we 
have constructed a diagnostic diagram whose properties have also been studied
in detail. In particular, the most important source of error in the determination of 
the mean density, and other quantities like the frequency of the fundamental
radial mode, $\nu_0$, comes from the uncertainty in 
the large spacing measure for rotating stars. In particular our results are 
valid only for stars rotating at most at 40\% of the break-up velocity.
Nevertheless, these uncertainties are expected to be drastically 
lowered by the inclusion of additional observational constraints (as shown by
GH09). 

Finally we have applied the diagnostic method presented in this work to five stars
for which regular patterns have been found. Our estimates for the mean density
and the frequency of the fundamental radial mode match reasonably (within 20\% of deviation)
with those given in the literature using asteroseismology. The comparison with other
independent (non-asteroseismic) methods to obtain the mean density of stars
reveals that asteroseismology provides, by now, the more precise estimates.

The present work implies a significant step to approach the asteroseismic studies 
of A-F, pulsating stars, to the level of precision achieved in solar-like stars. This is particularly 
relevant, not only to understand the structure and evolution of these stars but also for the study of 
planetary systems. Moreover, the observation of solar-like oscillations in star hosting planets provides a
significant help for the characterization of the planetary systems, through the 
precise knowledge of the mass, radius, mean density and age of
the host star \citep[e.g. ][]{JCD10}. The present work permits the extent of those 
capabilities to the study of planets orbiting around A-F pulsating stars. 
In fact, most of the  planets discovered  by direct imaging are found to be orbiting such
stars. These systems are critical to understand the spin-orbital interactions between the 
planets and the hosting star \citep{Winn09,Wright11}. For these, an estimate of the age
of the system is crucial to determine the mass of the planets discovered. Thanks
to the large amount of A-F stars observed by CoRoT and Kepler, the number of applications 
of the relations here derived open a new and important window to study the structure and 
evolution of these stars \citep{Uytt11,Moya10}.

In a forthcoming study, we examine the statistical properties of the frequency spectra by 
analysing their autocorrelation functions and by looking at the cumulative distribution functions 
of the frequency separations. Results on mode visibility \citep{Reese13} together with the
use of multi-colour photometry will help us to better asses how large separation is 
affected by rotation, and to improve the accuracy of the relation found in this paper.\\[0.1cm]

\begin{acknowledgements}
       JCS acknowledges support from the "Instituto de Astrof\'{\i}sica de Andaluc\'{\i}a (CSIC)" by
       the Spanish "Plan Nacional del Espacio" under projects AYA2012-39346-C02-01,
       and AYA2010-12030-E.  JCS also acknowledges support by the European project
       SpaceInn (FP7-SPACE-2012-1). AGH acknowledges support by the Spanish MINECO project  
       (ESP2004- 3855-C03-01) and by FCT (Portugal) grant SFRH/BPD/80619/2011. 
       ES and CR acknowledge support 
       from the Spanish Virtual Observatory financed by the Spanish Ministry of Science through grants 
       AYA2011-24052.  This research benefited from the computing resources provided by the European 
       Grid Infrastructure (EGI) and by the GRID-CSIC project (200450E494). 
\end{acknowledgements}

\bibliography{reflist_papers2_bibtex}
\bibliographystyle{aa}

\newpage

%--------------------------------------------------------------------------------------------------
\appendix
%--------------------------------------------------------------------------------------------------

%--------------------------------------------------------------------------------------------------
\section{Estimate of intrinsic errors of the method\label{apsec:errors}}
%--------------------------------------------------------------------------------------------------

In order to estimate the domain of validity of the method, it is necessary to determine
how sensitive is the distribution of $(\lsep,\rhom)_i$ points in Fig.~\ref{fig:figcontour} to 
variations of the physical parameters used to construct the grid of models.
%fig%
\begin{figure}[!ht]
   \centering
       \includegraphics[width=9cm]{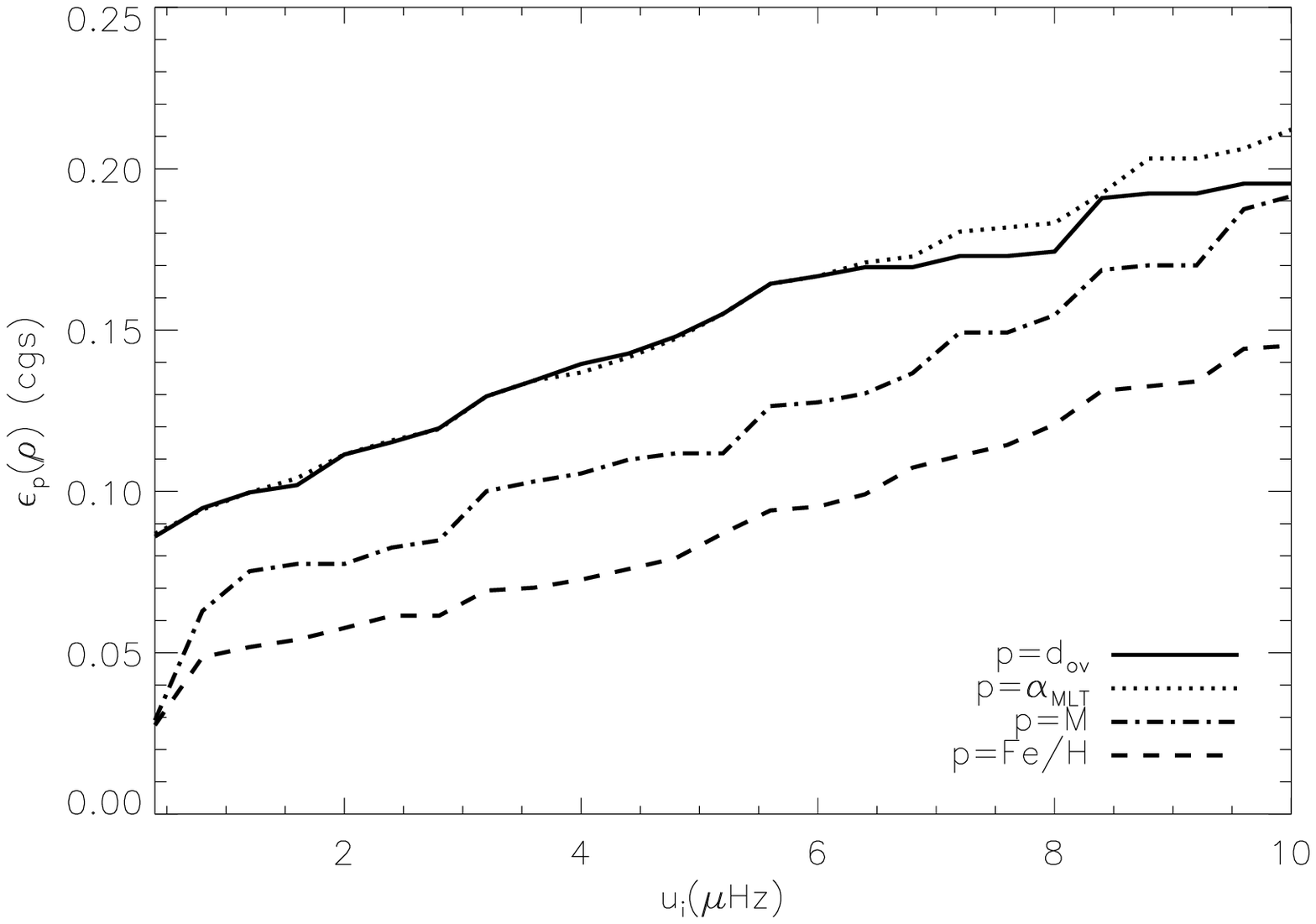}   
     \caption{Dependence of $\sigma_p(p)(\rhom)$ errors (Eq~\ref{eq:deferr}) with the uncertainty
                    in $\lsep$ for the four quantities considered in this work. A value of  $\lsep=60\,\muHz$ 
                    (middle of the main sequence approximately) is assumed. }
     \label{fig:errevol}
\end{figure}
%fig%
Notice that the linear fit given by Eq.~\ref{eq:linreg} was obtained considering
each model as an independent \emph{measurement} without uncertainties. Indeed, the whole dataset
is composed by independent subsets of models: the evolutionary tracks. As a consequence,
the standard procedure for calculation of coefficient errors and/or the goodness of the fit are 
meaningless here. Instead, we sought to estimate the errors committed by studying how variations
of the physical parameters with which the model dataset was constructed (Table~\ref{tab:param})
modify the shape or the thickness of the strip shown in Fig.~\ref{fig:figcontour}.

In particular we examined the $\lsep$-$\rhom$ relation by analyzing the maximum variation of a 
given parameter $p$ at once (Table~\ref{tab:param}), leaving the remaining parameters free to vary. 
Then, we calculated the size of the model strip at a given value of $\lsep_i$ with a given uncertainty 
$\pm u_i$, defined as
\eqn{S_{p}(\rhom) = |\rhom(\lsep_i+u_i) - \rhom(\lsep_i-u_i)|\,,\label{eq:defsize}}
and the intrinsic error committed in the estimate of $\rhom$ from Fig.~\ref{fig:figcontour} for a given
value $\lsep_i$ with an uncertainty of $u_i$ is given by
\eqn{\epsilon_{p}(\rhom) = {\rm max}\{S_{p_{\rm max}}(\rhom),S_{p_{\rm min}}(\rhom)\}\,.\label{eq:deferr}}
That is, we consider the maximum possible error for a given parameter variation.  

These error estimates are necessarily dependent on the uncertainty in the observed
value of the large spacings. 
We studied this dependence by calculating $\epsilon_p(x)$ for a
set of $u_i$ values (in $\muHz$) ranging from 0 to $10\,\muHz$. Figure~\ref{fig:errevol} shows
the evolution of the errors with $u_i$,  which is roughly linear. Notice that for 
$u_i\lesssim2\,\muHz$, values lower than $0.12\,\cgs$  are predicted for 
$\epsilon_p(\rhom)$ . Therefore, very low uncertainties 
in $\lsep$ are required for a good determination of the mean density. For instance, to get an 
uncertainty of $\pm0.02$ in $\rhom$, one would need to measure $\lsep$ with a 
precision $u_i$ lower than $1\,\muHz$. Note that this is the precision reached 
by \citet{GH13aa} in the study of periodicities of the \ds\ star \object{HD174966}.
Indeed, $\epsilon_p$ are intrinsic errors of the method used in this work. The total error 
$\sigma$  (see Table~\ref{tab:obs} on the mean density depend also on all the constraints
considered to model the studied star.
%fig%
\begin{figure*}[!ht]
   \centering
      \includegraphics[width=13cm]{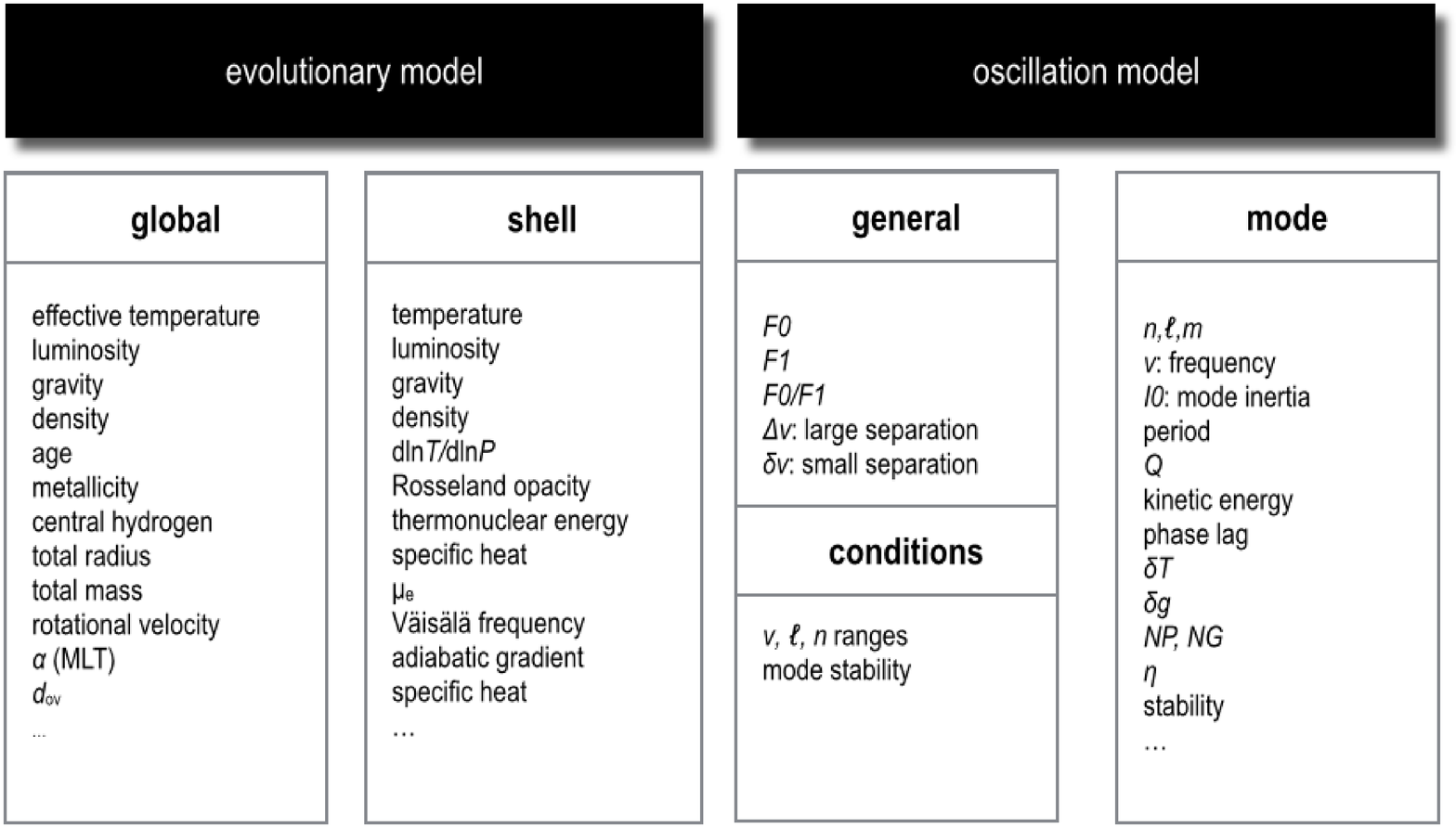} 
      \caption{Overall scheme of the basic data model adopted in \tool. The physical variables from
                       equilibrium models, and the asteroseismic variables are listed on the left, and right panels,
                       respectively.}
      \label{fig:dm}
\end{figure*}
%fig%

In the case of an ideal perfect measurement 
of the large spacing, the minimum precision is given by the maximum effect of the rotation
(due to the star deformation) on $\lsep$ (see Sect.~\ref{ssec:rolerot}). For very rapidly 
rotating objects (stars rotating faster than 40\% of the keplerian velocity) the effect of 
rotation on the large spacing is larger than $2.3\,\muHz$, which means  an intrinsic error 
on $\rhom$ of approximately $0.12 {\rm gcm}^{-3}$. For slower stars, this intrinsic error
becomes smaller. 
The maximum errors predicted for the estimate of the mean density range from 11\% to 21\%
of the total variation of $\rhom$ in the main sequence. 
 
 We recall that such variations correspond to the worst case, and therefore they must be regarded 
as an upper limit. When other observational constraints are considered (e.g., metallicity, gravity, 
effective temperature, etc.), the errors in the diagnostics here proposed can drop drastically, as 
it was shown in GH09.

%--------------------------------------------------------------------------------------------------
\section{\tool\ \label{apsec:tool}}
%--------------------------------------------------------------------------------------------------

Here we present \tool, the first virtual observatory tool for asteroseismology developed by the 
Spanish Virtual Observatory (SVO)\footnote{http://svo.cab.inta-csic.es}, with which has been
done the entire workflow of the study described in this paper. 
In this section we put the tool in context of necessities of the Asteroseismic 
community, describe its main objectives and characteristics, and detailed its
current workflow. Note that such an application is constantly evolving and
some of the snapshots here provided might be different in the future. In any 
case the main purpose and ultimate objectives will be kept.

%--------------------------------------------------------------------------------------------------
\subsection{Context \label{apsec:context}}
%--------------------------------------------------------------------------------------------------
Stellar physics experiments nowadays a significant progress thanks to the rapid development
of one of its main laboratories: the stellar seismology, which is the only technique allowing to 
probe the interior of stars for the detailed knowledge of the internal structure and the physical 
processes occurring there in. In the last decades we have witnessed a significant development 
of this technique, mainly thanks to
the increase of the quantity and quality of the observations particularly from space and
ground-based multisite campaigns. From space, a significant amount of high-quality asteroseismic data is available from: 
\most\ \footnote{http://www.astro.ubc.ca/MOST/}\citep[][]{most03}; 
\corot\ \footnote{Convection, Rotation, and planetary Transits. The CoRoT space mission was 
developed and is operated by the French space agency CNES, with participation of ESA's RSSD
and Science Programmes, Austria, Belgium, Brazil, Germany, and Spain: http://corot.oamp.fr/}\citep[][]{Corot03}, 
and \kepler\ \footnote{http://kepler.nasa.gov/} \citep[][]{Gilliland10}, launched in 2009.
Other missions scheduled on the near future like \gaia\ \citep[][]{gaia03} and 
\plato\ \citep[][]{plato09}, will increase by a factor of hundreds the available datasets. 
From the ground, dedicated photometric and spectroscopic follow-up observations for the above-mentioned
space missions \citep[e.g.][ for CoRoT and Kepler missions, respectively]{Poretti09,Uytt10an}
are necessary for the better characterization of the stars observed by the satellites.

A proper understanding of this huge amount of information requires a similar leap
forward on the theoretical side \citep[see ][ for a recent review on this topic]{Sua10lna}. 
Nowadays simulations of complex systems produce huge amounts of information that are difficult to
manipulate, analyze, extract and publish. Significant advances have been made on this
issue, e.g. the Asteroseismic Modeling Portal \citep[AMP, ][]{Metcalfe09} or MESA \citep{Paxton11} 
codes. However,  the main problem comes from the necessity of dealing with theoretical models 
developed by different groups, with different codes, numerical approximation, physical definitions,
etc. This lack of homogeneity makes it difficult to design automatic tools to simultaneously work with 
different models and/or applications able to use the models on the fly.

On the observational side, those problems have been successfully solved thanks to Virtual Observatory 
(hereafter, VO), which is an international initiative whose main objective is to guarantee an easy access 
and analysis of the information residing in
astronomical archives and services. Nineteen VO projects are now funded through national and
international programs, all projects working together under the IVOA\footnote{http://www.ivoa.net}
to share expertise and best practices and develop common standards and infrastructures for the VO.

In this context, the Spanish VO (SVO), which joined IVOA in 
June 2004, is deeply involved in the development of standards that guarantee a 
fully interoperability between theory and observations and among theoretical collections themselves.
In particular, SVO actively participated in the development of the VO access protocol for
theoretical spectra\footnote{http://ivoa.net/Documents/latest/SSA.html} and is presently working in
a more general protocol called S3\footnote{http://ivoa.net/Documents/latest/S3TheoreticalData.html}.
Examples of theoretical models published in the VO framework can be found at the SVO theoretical
model server\footnote{http://svo.cab.inta-csic.es/theory/db2vo4/}.
%fig%
\begin{figure}[!t]
   \centering
      \includegraphics[width=7.5cm]{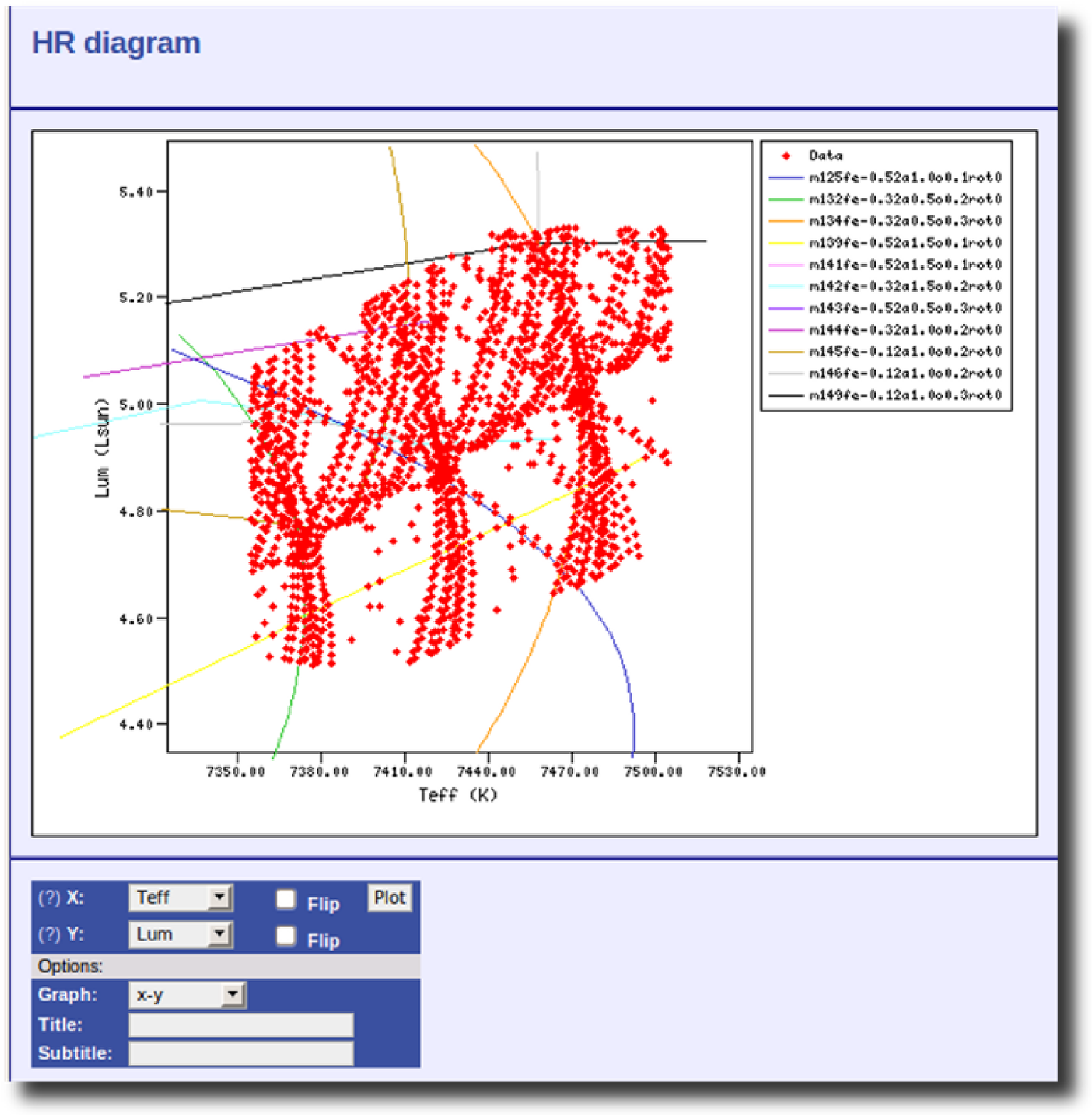}  
    \caption{Illustration of a Herzprung-Russel diagram of the models
                   matching all the input criteria simultaneously. Dots represent the effective
                   temperature and luminosity of all the valid models. Lines represent 
                   evolutionary tracks. For clarity, only some
                   of the models used in this work have been depicted. A colored version
                   of the plot is accessible in the online version of the paper.}
      \label{fig:hr}
\end{figure}
%fig%

%--------------------------------------------------------------------------------------------------
\subsection{Characteristics \label{apsec:charac}}
%--------------------------------------------------------------------------------------------------
\tool is a tool conceived to work with VO-compliant models. In the
Virtual Observatory, models are described according the same data model
and accessed using the same access protocol which solves all the issues
regarding data discovery, data access and data representation present in
non-VO tools.

The tool is intended to have a wider applicability in asteroseismology, and more generally in stellar 
physics and in any other field for which stellar models are required. To summarize, the main 
characteristics are:

\begin{itemize}
 \item[$\bullet$] Efficiency. \tool\ queries multiple model databases typically in seconds. 
 \item[$\bullet$] Collections of models are handled easily and with user-friendly web interfaces.
 \item[$\bullet$] The only software required is a web browser.
 \item[$\bullet$] Tables, figures, and model collections are fully downloadable.
 \item[$\bullet$] Designed for the easy and fast comparison of very different and heterogenous
       			    models.
 \item[$\bullet$] Visualization tools are available. Some of the plots presented in this work have been
                 	    built with the \tool\ graphic tools (see Appendix~\ref{apsec:workflow}).
 \item[$\bullet$] The tool offers new scientific potential, otherwise  technically impossible or
                 	    time consuming. The multivariable analysis performed in this work only required
                  	   a systematic query to \tool\ for the different parameters needed.  
\end{itemize}

Furthermore, \tool\ has also been designed for an easy and quick interpretation of the
asteroseismic data coming space missions, like \most, \corot, and \kepler, as
well as future missions like the PLAnetary Transits and Oscillations (\plato) , currently M3 candidate 
in ESA Cosmic Vision program, or the Transiting Exoplanet Survey Satellite (TESS), a new NASA 
space mission scheduled for launch in 2017. For this purpose, the next steps in the development of \tool\
will be:
\begin{itemize}
 \item[$\bullet$] The inclusion of new collections of models, namely solar-like and giant-like
                  asteroseismic models. This will be done by calculating new model datasets
                  with our own codes, and by adapting other model databases, built with differently
                  codes and different physics, to \tool.
 
 \item[$\bullet$] Implementation of direct link between \tool\ and other existing VO services, allowing
                  the search for observed physical parameters stored in VO-compliant databases
                  (namely those of the space missions), and using them as inputs in \tool.

\end{itemize}

%fig%
\begin{figure}[!t]
   \centering
     \includegraphics[width=9cm]{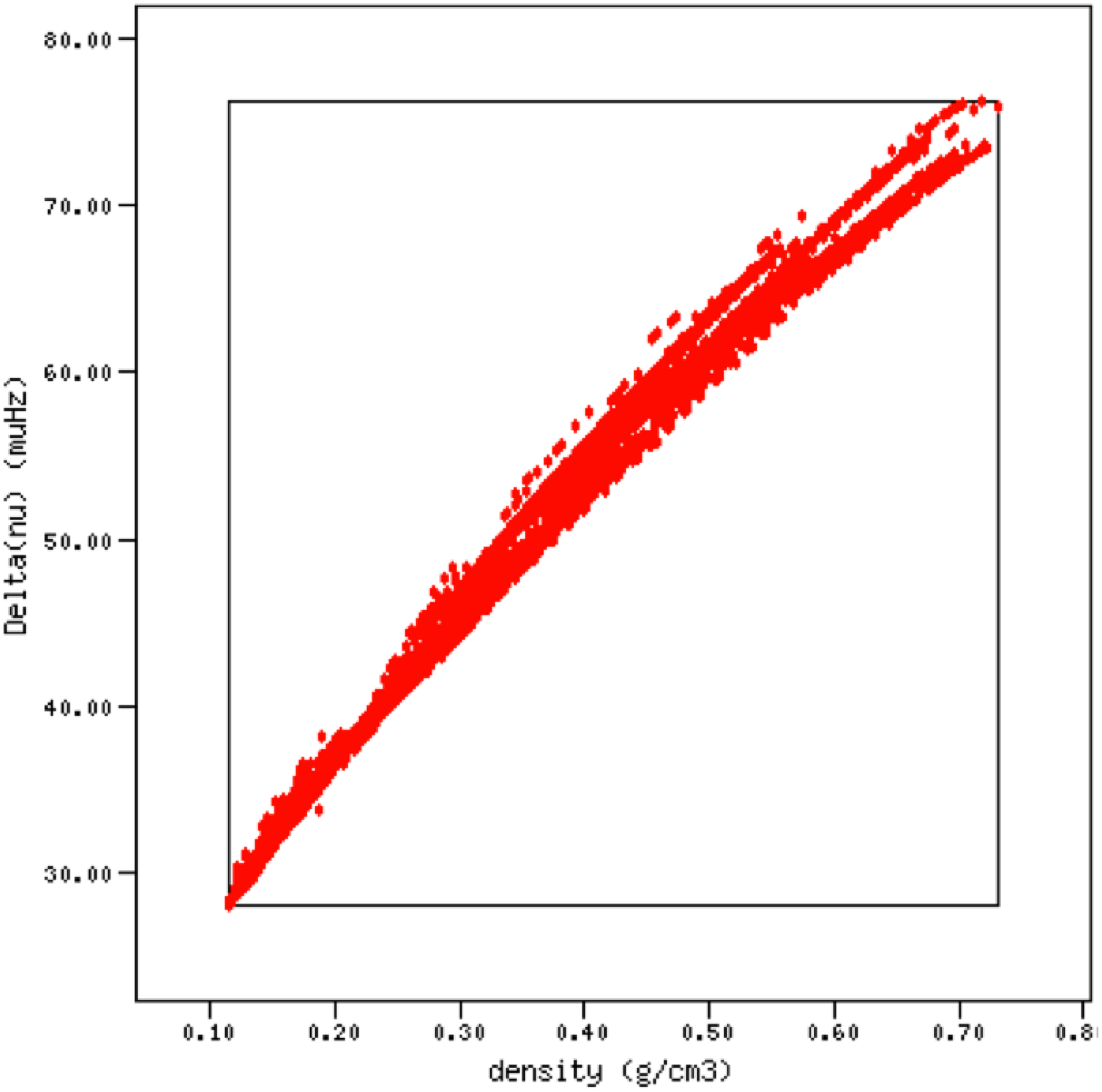}
     \caption{Large spacing, $\lsep$, as a function of the mean density for a large
                   set of models described in Sect.~\ref{ssec:ls-vs-ro}.  Large spacings 
                   were calculated using all the frequencies available per $\ell$, up to $\ell=3$
                   (more details in the text). Time evolution reads from right to left, as the mean 
                   density of stars in the main sequence decrease 
                   with time.  Plots were obtained using  \tool's 
                   graphical utilities. A colored version
                   of the plot is accessible in the online version of the paper.}
     \label{fig:gsdens}
\end{figure}
%fig%
%--------------------------------------------------------------------------------------------------
\subsection{VO service\label{apsec:voservice}}
%--------------------------------------------------------------------------------------------------
\tool\ has been designed following the Virtual Observatory standards and requirements. 
This means that, in parallel to the web interface, the system can also be accessed from other VO 
applications using the S3 protocol to obtain in a standard way information about:

\begin{itemize}
 \item The available combinations of evolutionary and seismological
           models.
 \item The query parameters, their physical description and the available range of values. 
 \item The list of models that match the query criteria and their properties.
 \item The stellar shell structure and the oscillation spectrum for each model.
\end{itemize}

From a technical point of view, this feature is very important, since it allows the tool to 
work with multiple model databases, no matter where they are physically located.
Moreover, this opens the possibility of interconnecting \tool\ with existing astronomical
archives, catalogs, etc., in particular those being constructed using asteroseismic space
data, already in VO-compliant form.

%--------------------------------------------------------------------------------------------------
\subsection{Workflow\label{apsec:workflow}}
%--------------------------------------------------------------------------------------------------

The \tool's workflow we describe here is general, and therefore applied for the present work.
It is composed of three main steps :
\begin{enumerate}
   \item[$\bullet$] Input parameters specification
   \item[$\bullet$] Summary of the results \& check out
   \item[$\bullet$] Model selection \& online analysis 
\end{enumerate}

One of the most critical steps when building a tool to handle different theoretical models in a
compatible way is the identification of the mandatory parameters to represent the physics involved,
and their mapping into a common set of variables. In this regard, we have developed a prototype data
model (Fig.\ref{fig:dm}) for asteroseismology which contains 17 star global properties (effective
temperature, surface gravity, luminosity, etc.), 44 star shell variables (density, pressure,
temperature, etc.), and 35 seismic properties (frequency ranges, fundamental radial mode, large and
small separation, etc.). For a maximum interoperability, we used the most common definitions
in the field for these variables, with the aim of setting the basis of VO standards for
asteroseismology. 

Once the model parameters has been selected, \tool\ queries the user-specified model database
Here we use our own model database described in Sect.~\ref{ssec:votadb}. The results obtained
from these queries are shown to the user in different formats, with the possibility of managing
them and, more importantly, of using \tool's online graphic tools which allow the researcher 
to easily do \emph{online asteroseismology}, for instance by

\begin{itemize}
     \item[$\bullet$]  Visually examining the resulting models, with some statistics and sorting
     				possibilities  for an efficient handling of the results.             
     \item [$\bullet$] Selecting individual or multiple files to be analyzed (including shell variables
     				analysis for equilibrium models) with the graphic tools, which allows the 
				user to download the generated plots (an example of HR diagram 
				build with \tool's graphic tools is shown in Fig.~\ref{fig:hr}).				             
      \item[$\bullet$] Allowing the user to select individual or multiple files to be downloaded  
      				(e.g. complete evolutionary tracks) in original codes' output formats and
				in VOTable formats. This provides compatibility with other VO-compliant
				visualization tools like TOPCAT\footnote{http://www.star.bris.ac.uk/~mbt/topcat/}.                  
     \item[$\bullet$] To download plots in "png" format, by placing the mouse pointer on the
              plot window and clicking the mouse's right button.
  
\end{itemize}

All these characteristics make it possible to easily perform quick \emph{on-the-fly} analysis 
of large set of models, and/or comparisons of different and heterogeneous models, or even
model collections.

Moreover, those tools allow the user to perform statistical works on theoretical properties 
of multiple variables at the same time. The present work is an example of such a work: the
 relation between the large spacings and the mean density for \dss\ shown in the contour
 plot (Fig.~\ref{fig:figcontour}) was built using the data obtained from Fig.~\ref{fig:gsdens} 
 during this research workflow.

\end{document}